\documentclass[12pt,preprint]{emulateapj}

\makeatother

\newcommand{\fgas}{$M_{\rm gas} / [M_{gas} + M_{\ast}]$}
\newcommand{\fg}{$f_{\rm gas}$}

\newcommand{\h}{{\it Herschel}}
\newcommand{\msol}{$\rm M_{\odot}$}

\newcommand{\lsol}{L$_{\odot}$}
\newcommand{\lir}{$L_{\rm IR}$}
\newcommand{\lco}{$L^{\prime}_{\rm CO}$}

\newcommand{\aco}{$\alpha_{\rm CO}$}

\newcommand{\mgas}{$M_{\rm gas}$}
\newcommand{\lam}{$\,\mu$m}
\newcommand{\cooz}{CO[J=1$\rightarrow$0]}
\newcommand{\cott}{CO[J=3$\rightarrow$2]}
\lefthead{Magdis et al..~}

\slugcomment{to appear in ApJ}
\shortauthors{Magdis et al.}
\begin{document}
 \title{The Molecular Gas Content of $z = 3$ Lyman Break Galaxies; Evidence of a non Evolving Gas Fraction in Main Sequence Galaxies at $z > 2$}
 \author{Georgios E. Magdis $\!$\altaffilmark{1},
 	     E. Daddi $\!$\altaffilmark{2},
 	     M. Sargent $\!$\altaffilmark{2},
		 D. Elbaz $\!$\altaffilmark{2},  
		 R. Gobat $\!$\altaffilmark{2},	 
		 H. Dannerbauer $\!$\altaffilmark{3},   
	     C. Feruglio $\!$\altaffilmark{4},
         Q. Tan $\!$\altaffilmark{2},
         D. Rigopoulou $\!$\altaffilmark{1,5},
         V. Charmandaris $\!$\altaffilmark{6,7,8},
         M. Dickinson $\!$\altaffilmark{9},
	     N. Reddy $\!$\altaffilmark{10},
	      H. Aussel $\!$\altaffilmark{2}
}          
\altaffiltext{1}{Department of Physics, University of Oxford, Keble Road, Oxford OX1 3RH}
\altaffiltext{2}{CEA, Laboratoire AIM, Irfu/SAp, F-91191 Gif-sur-Yvette, France}
\altaffiltext{3}{Universit\"at Wien, Institut \"ur Astronophysik, T\"urkenschanzstrasse 17, 1180 Wien, Austria}
\altaffiltext{4}{IRAM - Institut de RadioAstronomie Millim\'etrique 300 rue de la Piscine, Domaine Universitaire 38406 Saint Martin d'H\'eres, France}
\altaffiltext{5}{Space Science \& Technology Department, Rutherford Appleton Laboratory, Chilton, Didcot, Oxfordshire OX11 0QX}
\altaffiltext{6}{Department of Physics and Institute of Theoretical \& Computational Physics, University of Crete, GR-71003, Heraklion, Greece}
\altaffiltext{7}{IESL/Foundation for Research \& Technology-Hellas, GR-71110, Heraklion, Greece}
\altaffiltext{8}{Chercheur Associ\'e, Observatoire de Paris, F-75014,  Paris, France}
\altaffiltext{9}{NOAO, 950 N. Cherry Avenue, Tucson, AZ 85719, USA}
\altaffiltext{10}{Department of Physics and Astronomy
University of California, Riverside
900 University Avenue
Riverside, California 92507, USA}
\begin{abstract}
We present observations of the \cott ~emission towards two massive and infrared luminous  
Lyman Break Galaxies at $z = 3.21$ and $z = 2.92$, using the IRAM Plateau de Bure Interferometer, placing first constraints on the molecular gas masses (\mgas) of non-lensed LBGs.
Their overall properties are consistent with those of typical (Main-Sequence) galaxies at their redshifts, with specific star formation rates $\sim$ 1.6 and $\sim$ 2.2 Gyr$^{-1}$, despite their large infrared luminosities (\lir ~$\approx$ 2--3 $\times$ 10$^{12}$ \lsol) derived from \h.
With one plausible CO detection (spurious detection probability of $10^{-3}$) and one upper limit, we investigate the 
evolution of the molecular gas-to-stellar mass ratio (\mgas/$M_{\ast}$) with redshift. Our data suggest  that the steep evolution of \mgas/$M_{\ast}$ of normal galaxies up to $z \sim 2$ is followed by a flattening at higher redshifts,  providing supporting evidence for the existence of a  plateau in the evolution of the specific star formation rate at $z > 2.5$. 

\end{abstract}

%__________________________________________________________________
\section{INTRODUCTION}

Until recently, molecular gas observations in the distant Universe were restricted to very luminous, rare objects, while the gas content of normally star forming galaxies - that fall within the so-called ``Main Sequence'' relation between their star formation rates and stellar masses (e.g., Brinchmann et al. 2004; Noeske et al. 2007; Elbaz et al. 2011; Magdis et al. 2010a) - at $z > 2$ was unknown. However, in the last few years the first molecular gas surveys of normal, massive-star-forming galaxies at $z = 1-2$ were performed (Daddi et al. 2008; 2010; Tacconi et al. 2010), revealing that such typical distant star-forming galaxies were gas rich. Lyman Break Galaxies (LBGs) are the  most common population of normal star-forming galaxies at $z \sim 3$ (e.g., Steidel et al. 1999, Adelberger et al. 2000) and are ideal targets to extend this study to even higher redshifts. 

Multi-wavelength studies of LBGs have provided extensive information 
on various physical properties of these objects such as stellar masses and dust attenuation (e.g., Papovich et al. 2001, Rigopoulou et al. 2010, Magdis et al. 2008,2010a,b). However, their gaseous component still remains largely unconstrained, as currently the only two LBGs with molecular gas measurements are two gravitationally lensed systems: MS1512-cB58 (Yee et al. 1996) at $z = 2.72$ and  the LBG J213512.73-010143 --``the Cosmic Eye''--  (Smail et al. 2007), at $z = 3.07$. 
Baker et al. (2004), Coppin et al. (2007), and more recently Riechers et al. (2010) 
have reported CO detections for cB58 and Cosmic Eye, providing evidence of the existence of a sizeable cold gas reservoir in LBGs. The fact that these two galaxies are intrinsically faint and probably not representative of massive, $z \sim 3$ galaxies, coupled with the uncertainties introduced by lensing, implies that more data and direct CO observations of massive, normal high$-z$ galaxies are indispensable for an improved understanding of these systems. 

In this letter, we present CO measurements for two $z \sim 3$,  non-lensed LBGs in the  northern field of the Great Observatories Origins Deep survey (GOODS-N), selected to have properties representative of normal galaxies at this redshift. We couple these observations with data from the Herschel Space Observatory (Pilbratt et al. 2010), obtained as part of the GOODS-Herschel program (PI D. Elbaz), to study in detail their infrared properties and place constraints on their gas content and on the evolution of the gas to stellar mass (\mgas/$M_{\ast}$) evolution from $z = 2$ to $z = 3$. We assume $\Omega_{\rm m}$ = 0.3, H$_{0}$ = 71 km sec$^{-1}$ Mpc$^{-1}$, $\Omega_{\rm \lambda}$ = 0.7 and a Chabrier IMF.

%__________________________________________________________________

\section{SAMPLE AND OBSERVATIONS}
The two galaxies of this study (M23 and M18) were originally optically selected ($U_{n}, G, R$ with $ R < 25.5$) 
by Steidel et al. (2003) in the GOODS-N field. Rest-frame ultraviolet (UV) spectroscopy - which was also used to ascertain the absence of AGN signatures (i.e. strong high-ionization emission lines) - provided redshifts, $z_{spec} = 3.216$ and $z_{spec} = 2.929$, respectively, determined from Ly-$\alpha$  (among other) absorption lines.
The sources benefit from extensive multi-wavelength (UV to radio) coverage, including ground based $U_{n}, G, R, H, J, K$ observations and photometry from the Advance Camera for Surveys (ACS, $B,V, i, z$), Infrared Array Camera (IRAC) and the Multi-band Imaging Photometer (MIPS 24\lam) on board Spitzer and the Very Large Array (1.4\,GHz). The GOODS-{\it Herschel} program provides 100- and 160\lam ~imaging with the Photodetector Array Camera and Spectrometer (PACS, Poglitsch et al. 2010) and 250-, 350- and 500\lam ~with  the Spectral and Photometric Imaging Receiver (SPIRE, Griffin et al. 2010 ). \h ~photometry was 
performed using the source extraction point-spread function
(PSF) fitting code {\it galfit} (Peng et al. 2002), guided by 24\lam ~priors (for more details see Daddi et al. in prep). Both sources have a $>$ 3$\sigma$ detection in at least one \h ~band. Images of the two sources  in several bands, are shown in Fig. 1. 

We observed the CO[J=3$\rightarrow$2]  ($\nu_{\rm rest} = 345.796$\,GHz) transition towards the M23 and M18, using the Plateau de Bure Interferometer (PdBI). At $z = 3.216$ and $z = 2.929$, this line is redshifted to 82.0199 and 88.0112\,GHz respectively. 
Observations were carried out under good 3$\,$mm weather conditions in D array configuration with 5 antennas during May-June 2011,
resulting in 3.8h and 3.0h on source time after flagging of bad visibilities.  Data reduction was performed using CLIC in GILDAS. The noise level is 0.34 mJy/beam over 300 km s$^{-1}$ in both datasets, 
and the FWHM of the circularized synthetized beam is about 6$''$ for both.

\begin{figure}
\centering
\includegraphics[scale=0.29]{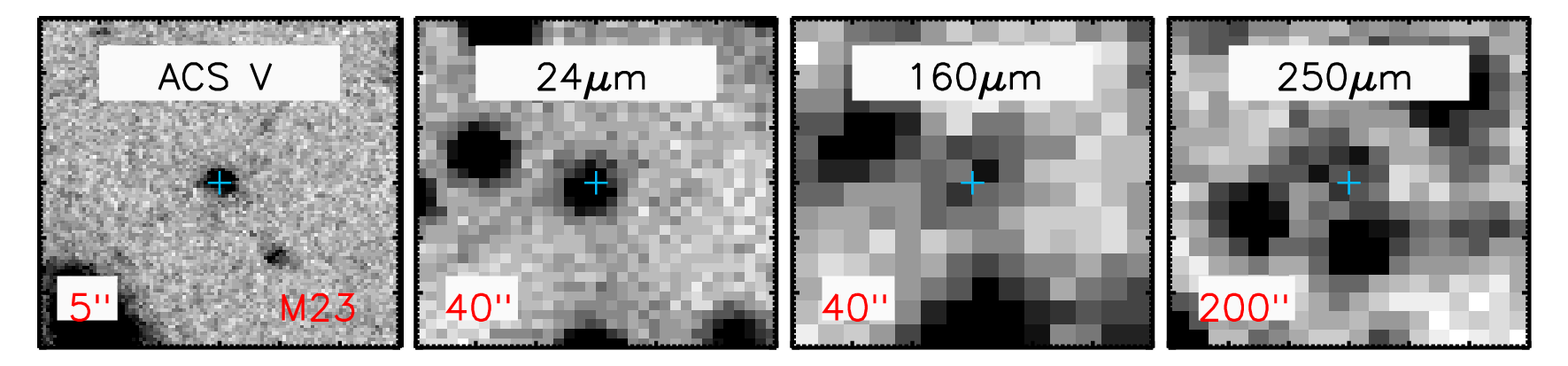}\\
\includegraphics[scale=0.29]{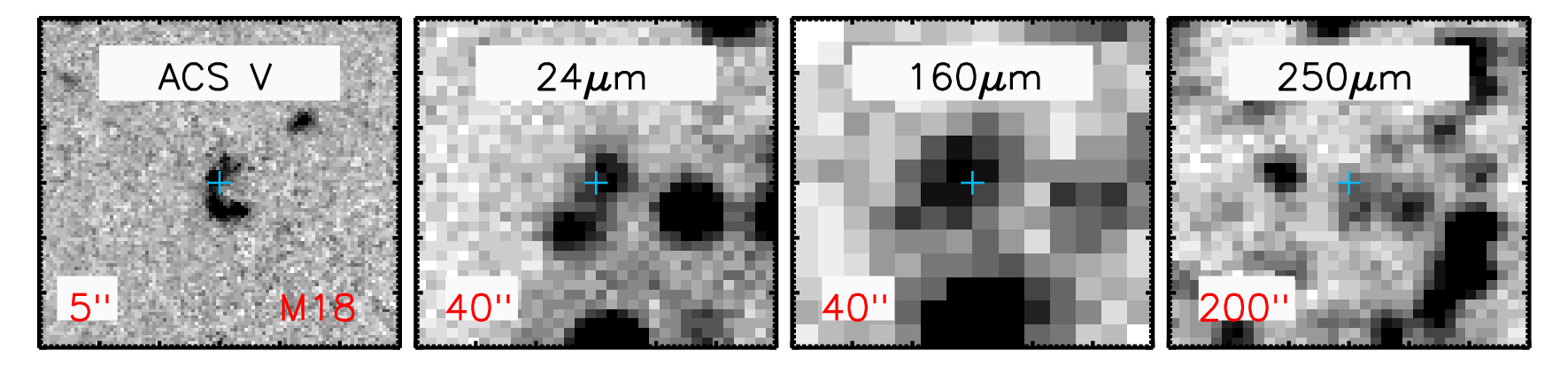}

\caption{ACS V-band (5''$\times$ 5''), MIPS 24\lam ~(40'' $\times$ 40''), PACS 160\lam ~(40'' $\times$ 40'') and SPIRE 250\lam ~(200'' $\times$ 200'')   cut-out images of M23 (top) amd M18 (bottom).
The cyan crosses are centred at the IRAC 3.6\lam ~position of the sources.}
\label{fig:sub} %
\end{figure}
\begin{figure*}
\centering
\includegraphics[scale=1]{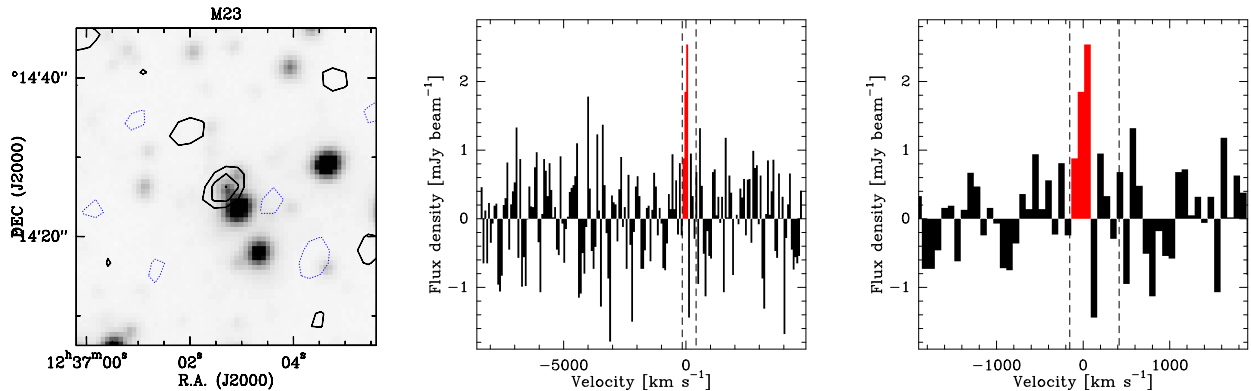}\\
\caption{CO observations of M23. left: Contours (from $\pm$ 2$\sigma$ and in steps of 1$\sigma$)  of CO[3-2] emission from M23 overlaid on IRAC 3.6$\,\mu$m images (40'' size). middle: Spectra of CO[3-2] emission binned in steps of 75 km s$^{-1}$ for the full range of velocity covered by our observations. right: Zoomed version of the middle panel, presenting only the 25\% of the total available spectral range. The red color indicates the regions where positive emission is detected. These regions have been used to derive total integrated fluxes. The vertical dashed lines indicate the velocity range that corresponds to the redshift uncertainty (1$\sigma$) as derived by optical spectroscopy (see Figure 3).}
\label{fig:sub} %
\end{figure*}

 \begin{figure*}
\centering
\includegraphics[scale=0.3]{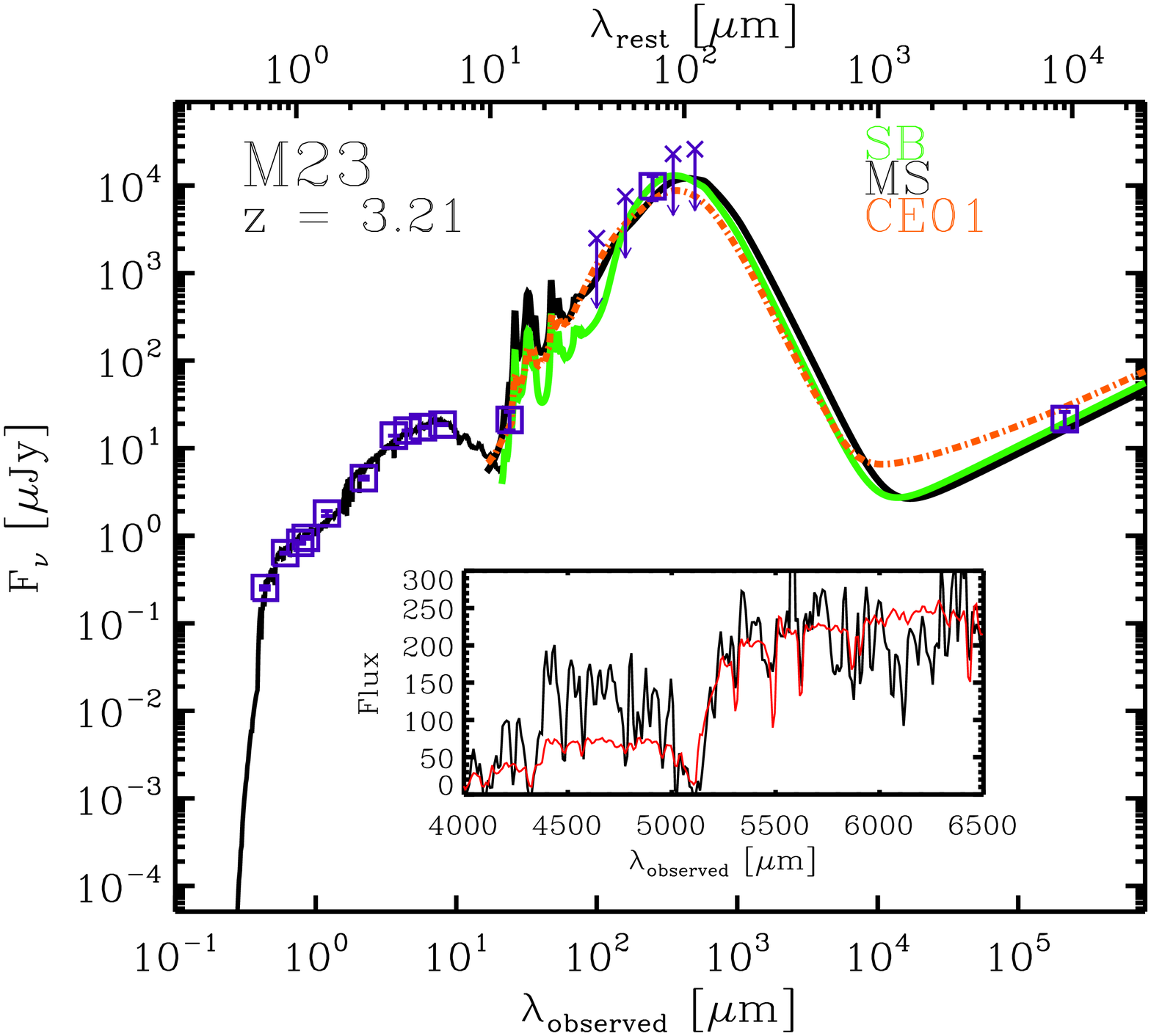}
\includegraphics[scale=0.3]{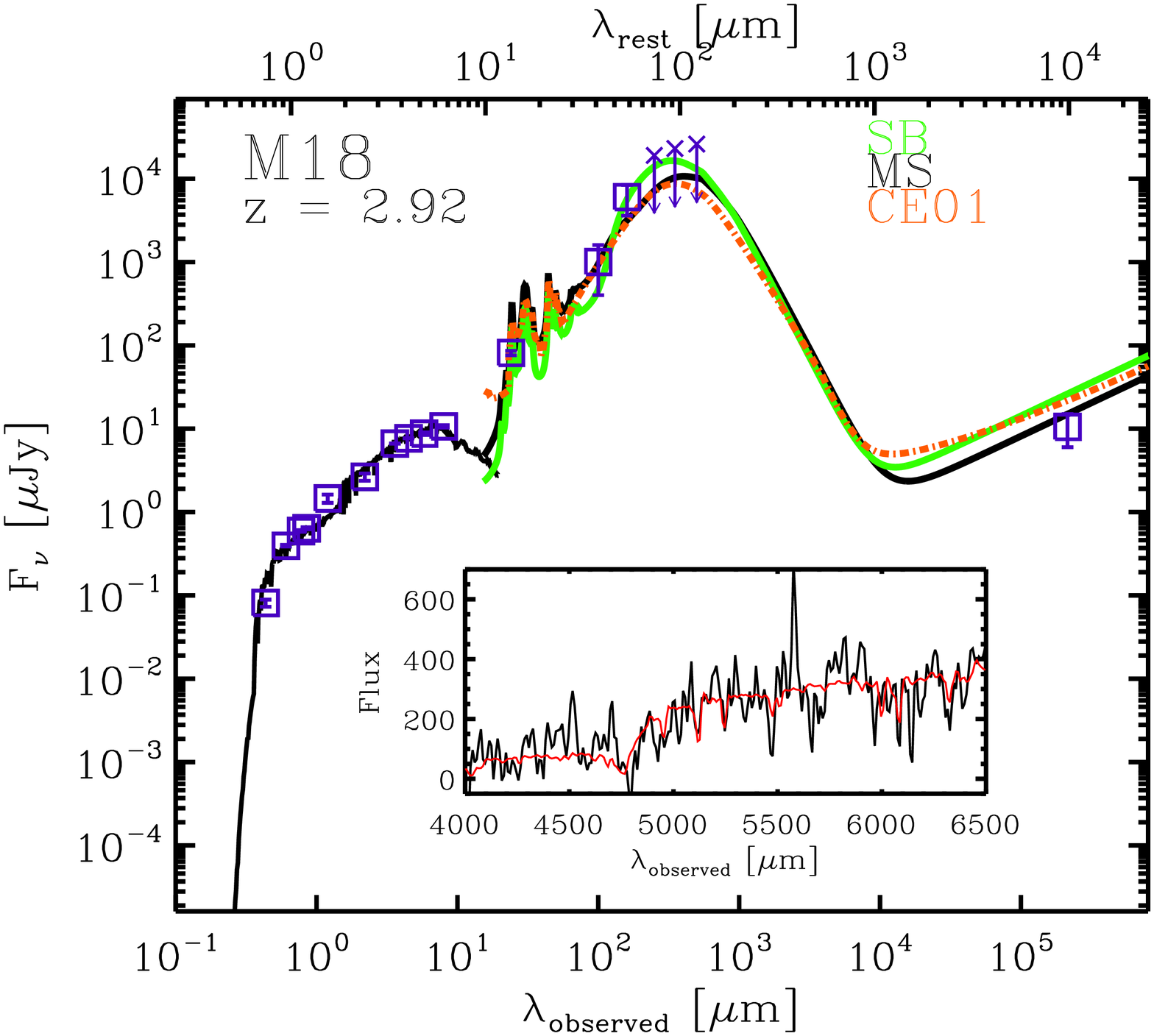}
\caption{Observed rest-frame UV to radio SEDs of M23 (left) and M18 (right). For the SEDs we present the detections (purple squares) along with 5$\sigma$ upper limits (purple arrows). The rest frame UV-near-IR portion of the data 
is overlaid with the best fit BC03 model (black solid line), while the mid-IR to radio is shown with the best fit 
CE01 model (orange solid line) and the best fit models of SB (green) and MS (black) templates from Magdis et al. (2011). Inset panels, Keck/LRIS rest-frame UV spectra of the sources (courtesy of A. Shapley). Red lines 
correspond to the best fit LBG spectral templates of Shapley et al. (2003).}
\label{fig:sub} %
\end{figure*}
\section{RESULTS}

We searched for CO emission at the position of the LBG targets (M23, RA: 189.2612457, DEC: 62.2405968 and M18, RA: 189.1837769, DEC: 62.2196693) and
 detected possible emission lines at  S/N = 4.0 and 3.3 for M23 and M18, respectively. The first is exactly at phase center while the latter reaches
3.6$\sigma$ if we allow for a small offset of 1.4$''$ (only 20\%
of the synthesized beam). In both cases this peak emission is within the velocity range allowed by the 1$\sigma$ optical redshift uncertainties of 600 km s$^{-1}$ and 1000 km s$^{-1}$ respectively, for M23 and M18. This is much smaller than the $>$12000 km s$^{-1}$ velocity range sampled by the 3.6 GHz WIDEX correlator. In order to quantitatively assess the likelihood of
chance/spurious detection of similar signals we extracted spectra from our datasets at 10000 fixed positions chosen within twice the primary beam but excluding the central LBG region. For each of this spectra we consider 600 km s$^{-1}$ (1000 km s$^{-1}$) subsets at a time for M23 (M18) and systematically look for maximally positive signals within this range. From this we estimate that the 
chance probability to have $S/N \geq 4.0$ for M23 is 1.4 $\times 10^{-3}$. Based on i) the fairly small chance of the signal being 
spurious, ii) the fact that the observed signal is spatially coincident with the position of the source (Fig. 2 left) and iii) that the emission is centred at the expected frequency within 1$\sigma$ of the optical spectroscopic uncertainty (Fig. 2 middle and right), we consider M23 to be a plausible detection, and derive a CO[3-2] luminosity of (1.60 $\pm$ 0.4) $\times$ 10$^{10}$ K km s$^{-1}$ pc$^{2}$. For M18, we only place a 4$\sigma$ upper limit of $<$ 2.2 $\times$ 10$^{10}$ K km s$^{-1}$ pc$^{2}$ (assuming a FWHM of 400 km s$^{-1}$), as the detected signal is weak, with a spurious probability of $3 \times 10^{-2}$.

\section{DISCUSSION}
\subsection{Physical Properties of the Galaxies}
We combine all available photometric data, ranging from rest-frame UV to observed radio 1.4GHz, to construct the full SEDs of the sources (Fig. 3). We fit  the infrared part of the SEDs using the Chary \& Elbaz (2001) templates as well as the 
Main-Sequence and starburst  SEDs, described in Magdis et al. (2011b), to derive estimates for the 
infrared luminosity of the sources. All templates result in similar estimates, indicating that both sources 
have ULIRG-like luminosities. In particular we find \lir ~=  (2.9 $\pm$ 1.6) $\times$ 10$^{12}$ \lsol ~ 
for M23  and  \lir ~=  (3.0 $\pm$ 1.1) $\times$ 10$^{12}$ \lsol ~for M18. Given the low S/N of the 
infrared data, we also employ the radio detections ($S_{1.4} = 20.1 \pm 5.2 \mu$Jy for M23 and 15.3 $\pm$ 4.8 $\mu$Jy for M18, Morrison et al. 2010) for an independent estimate of the \lir ~of the 
systems. In particular we convert the radio fluxes to $L_{\rm 1.4 GHz}$, assuming and slope of 
$\alpha = -0.8$ and based on the  Condon (1992) IR-radio correlation, we find \lir~= $4.8 \times 10^{12}$ \lsol ~for 
M23 and  \lir~= $1.9 \times 10^{12}$ \lsol ~for M18, in reasonable agreement with those derived based on 
SED fitting. 

Fitting the rest-frame UV-to-near-IR part of the spectrum
 with the Bruzual $\&$ Charlot (2003) model SEDs assuming a constant star formation history, yields a stellar mass
of $2.0 \times 10^{11}$ \msol ~for M23 and  $9.5 \times10^{10}$ \msol ~for M18. 
Based on the best fit models  the sources are found to be  moderately obscured with
similar extinction values ( $E(B-V)$ $\sim$0.25 ). Furthermore, using the extinction-corrected 
UV-luminosity $L_{\rm 1500}$ and adopting the Kennicutt (1998) relation we
derive dust corrected star formation rates that correspond to \lir ~$\approx$ 1.6 $\times$ 10$^{12}$ \lsol ~and  1.3 $\times$ 10$^{12}$ \lsol
~for M23 and M18 respectively. These UV based \lir ~estimates are $\sim$2 times lower compared to the  
 ``true'' \lir ~derived from the IR data. Nevertheless, this discrepancy is considerably lower than what is found for local ULIRGs and high$-z$ starbursts, and closer to that found for normal galaxies (e.g., Goldader et al. 2002, Elbaz et al. 2007, Magdis et al. 2010, Rigopoulou et al. 2010). 
 
Given that the far-IR, radio and UV \lir ~estimates agree within the uncertainties, we choose to average the three indicators and subsequently derive the best SFR measurements.  With SFR = 310 $M_{\odot}$ yr$^{-1}$ and 
210 $M_{\odot}$ yr$^{-1}$  for M23 and M18 respectively our analysis yields sSFR $\sim$ 1.6 Gyr$^{-1}$  and  $\sim$ 2.2 Gyr$^{-1}$ for the two sources. These values are very close to the characteristic, mass dependent sSFR$_{\rm MS}$ - derived based on the SFR-$M_{\ast}$ Main Sequence of $z \sim 3$ LBGs presented  by Magdis et al. (2010) - with sSFR/sSFR$_{\rm MS}$ $\sim$1.2 and 0.9 for M23 and M18 respectively. Consequently, despite the high SFRs of our galaxies that are much higher than those of most common $L_{*}$ galaxies at $z\sim3$, their overall properties are consistent with those of normal rather than to those of starburst galaxies that tend to have substantially elevated SFRs for their stellar masses. 
%We note that the sSFR, using the far-IR based \lir ~would be 1.5 Gyr$^{-1}$ for M23 and 3.1 Gyr$^{-1}$ for M18.
\begin{table*}
{\scriptsize
\caption{Observed and Derived Properties}             
\label{tab:4}      
\centering                          
\begin{tabular}{l c c c c c c c c c c c c}        
\hline\hline                 
Source &$z^{a}$ & $M_{\ast}$ & \lir $^{b}$& $I_{\rm CO[3-2]}$&S/N det&$ L_{\rm CO}$ $^{c}$ & \mgas $^{d}$ &\fg \\

& &$M_\odot$ &\lsol &Jy km s$^{-1}$&&K~km~s$^{-1}$~pc$^{2}$ &  $M_\odot$&\\
\hline
M23 &3.216& (2.0$\pm$0.3) $\times$ 10$^{11}$ &(3.1$\pm$1.1) $\times$ 10$^{12}$& 0.36$\pm$0.08 & 4.0 &~~(3.20$\pm$0.8) $\times$ 10$^{10}$ & (1.15$\pm$0.3)  $\times$10$^{11}$ & 0.36\\ 
M18 &2.929& (9.5$\pm$1.4) $\times$ 10$^{10}$ &(2.1$\pm$1.3) $\times$ 10$^{12}$& $<$0.52&  - &$<$4.40 $\times$ 10$^{10}$ &$<$1.58  $\times$10$^{11}$&$<$0.62\\ 
\hline                                   
\end{tabular}\\
Notes:\\
a: $\Delta(z)$ = 0.004 and 0.006 for M23 and M18 respectively\\
b: averaged \lir ~from UV, far-IR and radio estimators \\
c: from CO[3-2] emission line, derived after applying $r_{\rm 31}$ = 0.5 \\
d: assuming \aco=3.6, including He\\ }
\end{table*}

 \subsection{Molecular Gas Mass}
CO measurements can be used to derive estimates of the molecular gas mass of a galaxy (e.g., Solomon \& Vanden Bout 2005). 
However, the conversion factor from CO luminosities to the molecular gas mass,  \mgas~= \aco $\times$ \lco, introduces a large uncertainty in this derivation. In particular Downes \& Solomon (1998) showed that \aco ~is a factor of $\sim$ 6 smaller for local ULIRGs, than for local spiral galaxies with average values of ~0.8 (but also see Papadopoulos et al. 2012) and 4.6, respectively. Quite naturally, it has been a common practice to apply the same \aco ~value for both local and high$-z$ ULIRGs. Nevertheless, recently, there has been substantial evidence (both theoretical and observational),  that the \aco ~value for high$-z$ Main Sequence galaxies, is comparable to that of local spirals and that it depends on the metallicity rather than the infrared luminosity of the source (e.g., Magdis et al. 2011, Genzel et al. 2011, Narayanan et al. 2011, Daddi et al. 2010). In light of these results, and given that our LBGs are Main Sequence galaxies, we will attempt to make an educated guess of the \aco ~value for our LBGs, based on the \aco ~$-$ metallicity relation for normal galaxies (Genzel et al. 2011). 

Since we lack direct measurements of the metallicity for the two sources, we employ the fundamental 
metallicity relation of Mannucci et al. (2010) that relates the SFR and the stellar mass to metallicity. We derive a nearly solar metallicity for both sources, (i.e., $12+$log$(O/H)=8.7$) that based on the Genzel et al. (2011) relation would give \aco\footnote{The units of \aco, \msol pc$^{-2}$ (K km s$^{-1}$)$^{-1}$, are omitted from the text for brevity. The quoted values include helium.} ~$\sim$ 3.6. A similar value would be derived based on the local \aco ~- $Z$ relation of Leroy et al. (2011). We also note that this value is very close to the one derived by Magdis et al. (2011) and Daddi et al. (2010), for normal BzK galaxies, using  dust mass measurements and dynamical constraints, respectively. We convert luminosities derived from \cott ~to \lco ~(defined relative to the fundamental \cooz ~transition),  adopting an excitation correction of $r_{\rm 31}=0.5$ based on the CO spectral line energy distribution measured by Dannerbauer et al. (2009) for a BzK galaxy at $z \sim1.5$. Then based on the above we derive \mgas ~= (1.15 $\pm$ 0.30) $\times$ 10$^{11}$ $\times$ (\aco/3.6) \msol ~for M23 and an upper limit of  $<$ 1.58 $\times$ 10$^{11}$ $\times$ (\aco/3.6) \msol ~for M18. 

Focusing on M23, we infer a gas consumption timescale ($\tau_{\rm gas}$ = \mgas/SFR) of $\sim$ 0.4 Gyr, very similar to that of normal disks at $z \sim1$-2 by Daddi et al. (2010) and Genzel et al. (2010) and a factor of $\sim$ 7, higher than that of local ULIRGs. This suggests that it can maintain its star formation activity for much longer time scales, than those expected for rapid, merger-driven bursts, highlighting a fundamental difference in the star-formation mode of local ULIRGs and high$-z$ Main Sequence galaxies with comparable infrared luminosities. We note that although this result is subject to several uncertainties, it is unlikely to be driven by systematics in the adopted \aco ~value. By simply examining at direct observables, \lco ~and \lir, M23 has a star formation efficiency (SFE = \lir/\lco) $\sim$ 95 \lsol~(K km s$^{-1}$ pc$^{2}$)$^{-1}$, similar to that of high$-$z disks (Daddi et al. 2010). This is considerably lower when compared to the two lensed LBGs in the literature for which CO observations are available, the Cosmic Eye and the cB58 (Riechers et al. 2010), with \lir/\lco $\sim$260 and $\sim$600 \lsol~(K km s$^{-1}$ pc$^{2}$)$^{-1}$ respectively. Furthermore, the two lensed LBGs, exhibit an excess in their sSFR by a factor of $\sim$6 with respect to Main Sequence galaxies at this redshift (Riechers et al. 2010). The elevated sSFR and the high SFE of the lensed LBGs, typical characteristics of  merger driven star-bursting systems (e.g., Daddi et al. 2010, Elbaz et al. 2011), are in striking contrast with the ``normal'' star forming mode of M23.  Evidently a larger sample of non-lensed $z \sim 3$ LBGs is essential to drive this  investigation.
\begin{figure*}
\centering
\includegraphics[scale=0.6]{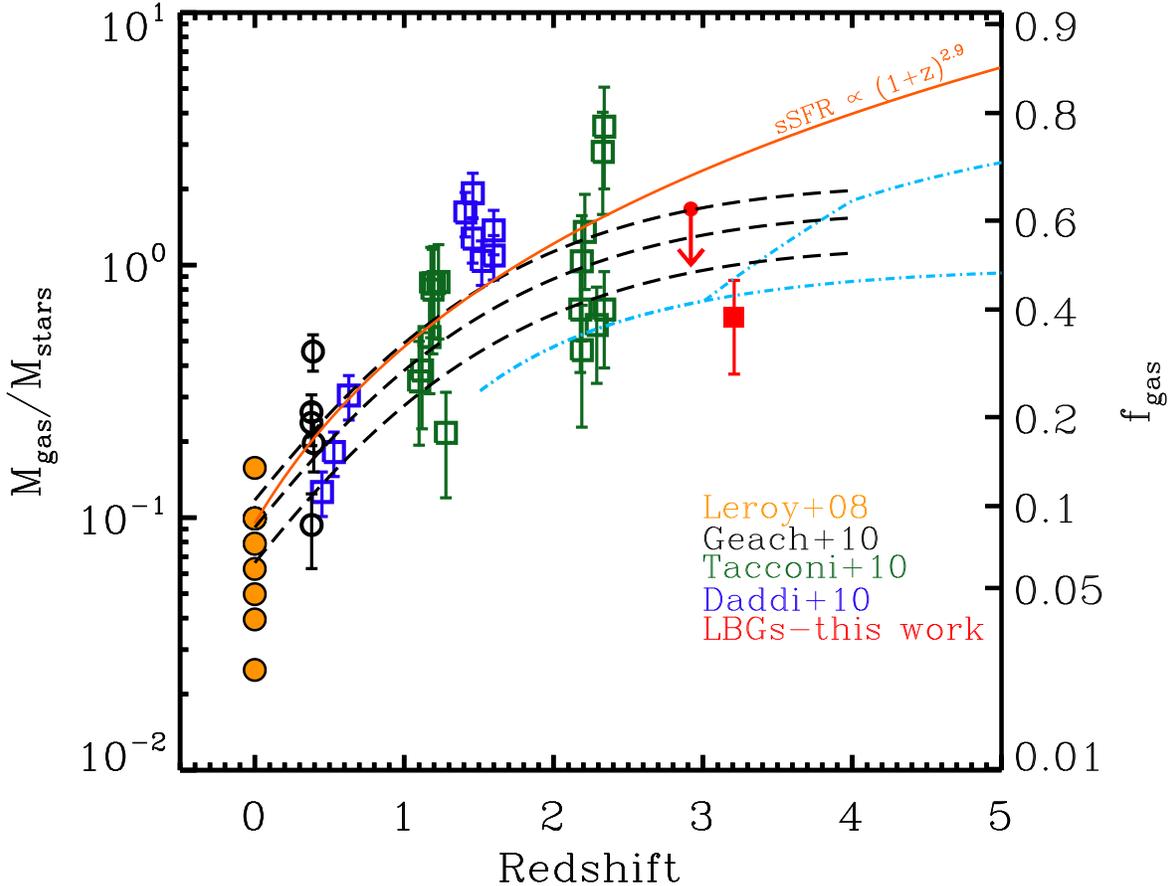}
\caption{Evolution of  \mgas/$M_{\ast}$ and \fg ~with redshift. Literature data are: 
orange circles from Leroy et al. (2008), empty black circles from Geach et al. (2011), 
blue squares from Daddi et al. (2011) and Salmi et al. (2012) in prep., green squares from 
Tacconi et al. (2010). The $z \sim 3$ LBGs of this study are shown with filled red squares. M18 is presented as a 4$\sigma$ upper limit. Dashed lines depict the tracks of Sargent et al. (2012b in prep), for the case of $M_{\ast}$ = 5.0 $\times$ 10$^{10}$ \msol, $M_{\ast}$ = 1.0 $\times$ 10$^{11}$ \msol ~and $M_{\ast}$ = 2.0 $\times$ 10$^{11}$ \msol. Cyan lines show the evolutionary tracks of Reddy et al. (2012), for galaxies with a final stellar mass  $M_{\ast}$ = 5.0 $\times$ 10$^{10}$ \msol ~at $z = 2.3$ for the case of a  constant sSFR = 2.4 Gyr$^{-1}$ at $z \geq 2$ and for the case of sSFR = 2.4 Gyr$^{-1}$  up to $z \sim 3.0$  and sSFR = 5.0 Gyr$^{-1}$ at $z \sim$ 4. The solid orange line depicts the predicted 
evolution of  \mgas/$M_{\ast}$ for the case of an increasing sSFR $\propto (1+z)^{2.9}$ at all redshifts (Dav\'e et al. 2012).}
\label{fig:sub} %
\end{figure*}
\subsection{The evolution of the Gas-to-Stellar Mass Ratio}
Currently, one of the most notable debates between theoretical predictions and observational evidence, is the evolution of sSFR beyond $z \sim 2$. In particular, several 
observational studies (but not all, see de Barros et al. 2012 for example) indicate that the sSFR evolves steeply up to $z\sim$2, and then remains 
roughly constant throughout the redshift range $z = 2-7$ (e.g., Gonz\'alez et al. 2010; Daddi et al.\ 2010). On the other hand, current theoretical approaches fail to reproduce this plateau, predicting a gradual increase of sSFR at higher redshifts (e.g., Dekel et al. 2010,  Weinmann et al. 2011, Dav\'e, et al. 2012). Given that the gas mass of normal galaxies is intimately linked with their star formation with \mgas ~$\propto$ SFR$^{0.81}$ at all redshifts (e.g., Daddi et al.\ 2010), the evolution of gas-to-stellar mass ratio (\mgas/$M_{\ast}$) should be  closely linked to the evolution of the sSFR, albeit with a shallower slope. Indeed, since sSFR=SFR/$M_{\ast}$, it implies that at a fixed stellar mass, \mgas/$M_{\ast} \propto$ sSFR$^{0.81}$. Therefore, \mgas ~estimates can be used to trace the evolution of sSFR in an independent way.

While restricted to only two galaxies,  we can place constraints on the evolution of  \mgas/$M_{\ast}$, or 
equally of the molecular gas fraction (\fg~= \fgas),  of normal galaxies up to $z \sim 3$ by  combining our data with results from previous studies of normal galaxies at various redshifts. In Figure 4, we present estimates of  \mgas/$M_{\ast}$, for normal 
galaxies in the local universe from Leroy et al. (2008), at $z \sim 0.4$ from Geach et al. (2011), at $z \sim 0.5$ from Salmi et al. (2012) (in prep.), and at $z 
\sim1-2$ from Daddi et al. (2010) and Tacconi et al. (2010), including only sources with $M_{\ast}$ $\ge$ 10$^{10}$ \msol. For our LBGs we find that  \mgas/$M_{\ast} \approx 0.6$ (M23) and \mgas/$M_{\ast} < 1.6$ (M18).

To test the two competing scenarios regarding the evolution of sSFR at $z > 2.5$,  we plot in Figure 4 the evolution of \mgas/$M_{\ast}$ for the case of a gradually increasing sSFR as $(1+z)^{2.9}$ at all redshifts, driven by the dark matter specific accretion rate as predicted by the theoretical work of Dav\'e et al.\ (2012). We also plot  the observationally motivated tracks of Sargent et al. (2012b in prep), that are built assuming \mgas ~$\propto$ SFR$^{0.81}$ at all redshifts and an increase of sSFR  as $(1+z)^{2.8}$ up to $z \sim 2.5$ followed by a flattening towards higher redshifts, as well as the evolutionary tracks of Reddy et al. (2012) for $L_{\ast}$ galaxies at $z \sim 2-7$. Our \mgas/$M_{\ast}$ estimate and upper limit, appear to be consistent with a flattening of the evolution, providing supporting evidence for a plateau in the evolution of sSFR too, at $z > 2.5$.  
We note that  an anti-correlation between \mgas/$M_{\ast}$ ~and $M_{\ast}$ is expected given that (i) the sSFR is not mass-invariant but declines with $M_{\ast}$ and (ii) the relation 
between \mgas\ and SFR is not linear. Therefore, the fact the our LBGs are 2 to 4 times more massive than the comparison sample (that has $\langle M_{\ast} \rangle \approx 5 \times 10^{10} M_{\odot}$) could introduce a bias in our 
result. However, rescaling the  \mgas/$M_{\ast}$ of all galaxies in Figure 4 to  $M_{\ast}$ = 5.0 $\times$ 10$^{10}$ \msol ~by assuming  \mgas/$M_{\ast} \propto M^{-0.4}_{\ast}$ (e.g., Daddi et al. 2010, Saintonge et al. 
2011, Magdis et al. 2012, submitted), does not affect our conclusion. Finally, the flattening of the evolution would be even  more pronounced if the observed signal from both of our  sources were treated only as upper 
limits and is further observationally supported by the non detection of CO emission from $z \sim 5$ LBGs (Davies et al.\ 2010).

Based on observations carried out with the IRAM Plateau de Bure Interferometer. IRAM is supported by INSU/CNRS (France), MPG (Germany) and IGN (Spain). Based also on observations carried out by the Herschel space observatory. Herschel is an ESA space observatory with science instruments provided by European-led Principal Investigator consortia and with important participation from NASA.
We thank an anonymous referee whose remarks helped clarify several aspects of this paper.
We are grateful to Alice Shapley for providing the optical spectra of the sources. 
GEM acknowledges support from the University of Oxford and the Fell Fund.
ED, MB, ED and MTS were supported by grants 
ERC- StG UPGAL 240039 and ANR-08-JCJC-0008.VC also acknowledges partial support by the COST Action ECOST-STSM-MP0905-230512-016069.

\clearpage

\end{document}